\documentclass[
reprint,superscriptaddress,amsmath,amssymb,aip]{revtex4-2}
\usepackage[multiple]{footmisc}
\usepackage[utf8]{inputenc}
\usepackage{amsmath,amsthm,amssymb,amsfonts,braket,mathtools}
\usepackage{placeins}[verbose]
\usepackage{tabularx, longtable, multirow}
\usepackage{graphicx}
\usepackage{bm,dsfont}
\usepackage[english]{babel}
\usepackage{comment}
\usepackage[dvipsnames]{xcolor}

\begin{document}

\title{Correcting Aberrations of a Transverse-Field Neutron Resonance Spin Echo Instrument}
\author{Stephen J. Kuhn}
\email{stkuhn@iu.edu}
\affiliation{Department of Physics, Indiana University, Bloomington, IN 47408, USA}

\author{Sam McKay}
\affiliation{Department of Physics, Indiana University, Bloomington, IN 47408, USA}
\affiliation{Quantum Science and Engineering Center, Indiana University, Bloomington, IN 47408, USA}

\author{Fankang Li}
\affiliation{Neutron Sciences Directorate, Oak Ridge National Laboratory, Oak Ridge, Tennessee 37830, USA}

\author{Robert M. Dalgliesh}
\affiliation{ISIS, Rutherford Appleton Laboratory, Chilton, Oxfordshire, UK}

\author{Eric Dees}
\affiliation{Department of Physics, Indiana University, Bloomington, IN 47408, USA}

\author{Kaleb Burrage}
\affiliation{Neutron Sciences Directorate, Oak Ridge National Laboratory, Oak Ridge, Tennessee 37830, USA}

\author{Jiazhou Shen}
\altaffiliation{Current address: Paul Scherrer Institut, Villigen, Switzerland}
\affiliation{Department of Physics, Indiana University, Bloomington, IN 47408, USA}

\author{Roger Pynn}
\affiliation{Department of Physics, Indiana University, Bloomington, IN 47408, USA}
\affiliation{Quantum Science and Engineering Center, Indiana University, Bloomington, IN 47408, USA}
\affiliation{Neutron Sciences Directorate, Oak Ridge National Laboratory, Oak Ridge, Tennessee 37830, USA}

\date{\today}
\begin{abstract}
The neutron resonance spin echo (NRSE) technique has the potential to increase the Fourier time and energy resolution in neutron scattering by using radio-frequency (rf) neutron spin-flippers. However, aberrations arising from variations in the neutron path length between the rf flippers reduce the polarization. Here, we develop and test a transverse static-field magnet, a series of which are placed between the rf flippers, to correct for these aberrations.The prototype correction magnet was both simulated in an NRSE beamline using McStas, a Monte Carlo neutron ray-tracing software package, and measured using neutrons. The results from the prototype demonstrate that this static-field design corrects for transverse-field NRSE aberrations.
\end{abstract}

\maketitle

\section{Introduction} 

\begin{figure*}
\centering
\includegraphics[width=.95\textwidth]{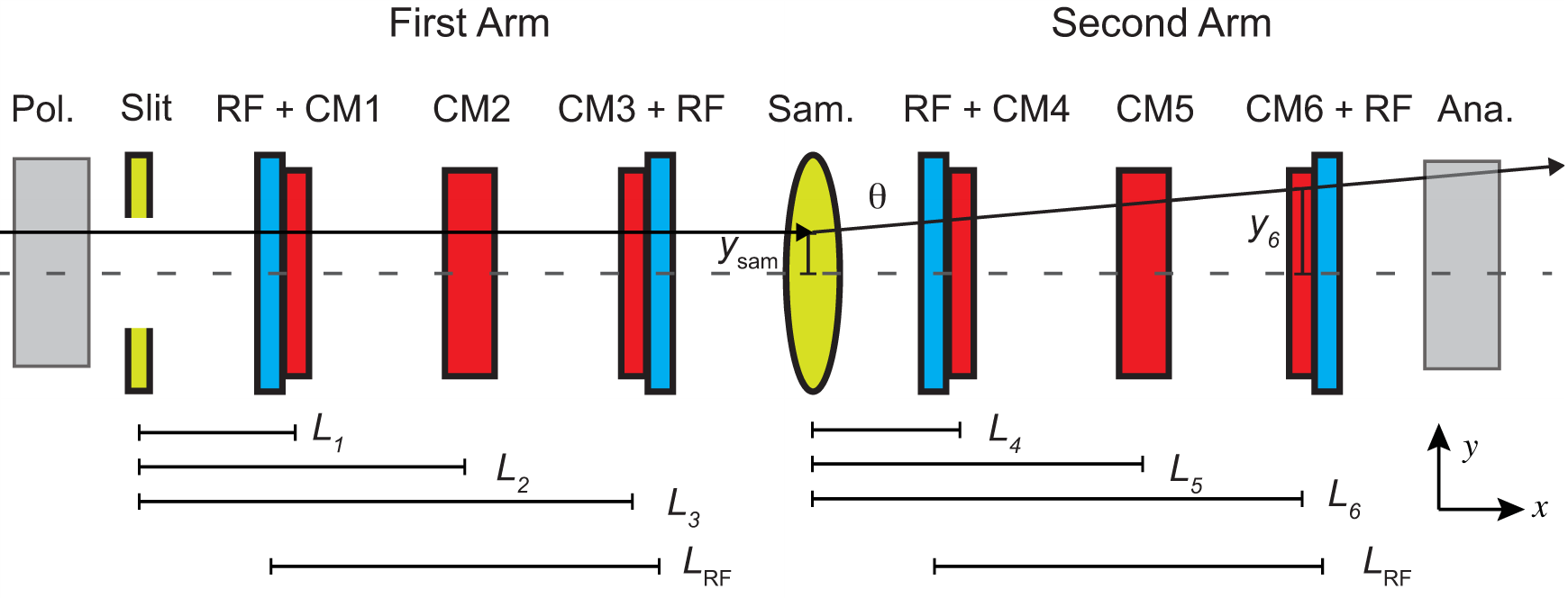}
\caption{ \label{figSchematic} Schematic of the corrected neutron resonance spin echo beamline. The neutron is polarized at the far left (Pol.) before traveling through the first arm, scattering from the sample (Sam.) at angle $\theta$, echoing through the second arm, and entering the analyzer (Ana.) at the far right; the detector is not shown. The correction magnets are labeled ``CM'' and the rf flippers ``RF''. The white space between the CMs is at zero field. The scattering plane for the example neutron path shown is the $x$-$y$ plane.
For the first arm, $L_n$ for $n=1,2,3$ is the distance from the beam-defining slit to the center of the corresponding correction magnet, while for the second arm, $L_n$ for $n = 4,5,6$ is the distance from the sample to each correction magnet center. The distance from the optical axis to the point that the neutron passes through the $n^{\mathrm{th}}$ correction magnet is defined as $y_n$. The distance between rf flippers in both arms is $L_{\mathrm{RF}}$.
}
\end{figure*}

Neutron Resonance Spin Echo (NRSE) is a modification of the Neutron Spin Echo (NSE) technique which replaces the static-field precession coils with radio-frequency (rf) spin-flippers.  \cite{Golub1987} The underlying principle of both types of \textit{echo} measurements is that the instrument will measure the change in velocity, and hence the change in energy, of a neutron scattered from a sample.
Currently, most large neutron sources use static-field NSE instruments for high-energy-resolution measurements of slow dynamics.

In order to be competitive with existing NSE instruments, NRSE would need to achieve a Fourier time (also called the spin echo time) of about one hundred nanoseconds. \cite{Farago2015,Ohl2012} 
The Fourier time $\tau$ is given by
\begin{equation}
    \tau = \frac{2 m^2}{h^2} L_{\mathrm{RF}} f \lambda^3
    \label{EqnFouriertime}
\end{equation}
where $L_{\mathrm{RF}}$ is the distance between the rf flippers in each arm, $f$ is the rf flipper (linear) frequency, $\lambda$ is the neutron wavelength, $m$ is the neutron mass, and $h$ is Planck's constant.\cite{Keller2002}
State-of-the-art rf flippers are already capable of producing a high performance NRSE instrument. As an example, suppose an NRSE beamline uses recently developed transverse rf flippers. \cite{Li2020} A beamline with those rf flippers operating at 4 MHz, with a 2 meter separation between the rf flippers, and with a standard NSE wavelength of 1 nm would have a Fourier time of 100 nanoseconds, which is comparable to modern NSE beamlines.\cite{Golub1988}

However, due to the long wavelength requirements for both NSE and NRSE, the neutron flux is often low and the relevant samples scatter weakly. Therefore, measurements are only possible by having a large spatial and angular beam size, which leads to aberrations. In conventional NSE, one aberration source is due to the variation in the static field strength across the beam due to the field profile created by a solenoid geometry.
In NRSE, the rf flippers are separated by zero field regions, so this aberration is not present.  A second type of aberration arises from scattering from a sample. The sample is placed in the center of the two symmetric arms as shown in Fig. \ref{figSchematic}. If the neutron scatters with some non-zero momentum transfer, then the path length through the second arm will not be the same as the path length through the first arm, so the neutron will spend a different amount of time in the two arms.
Because NRSE instruments measure the velocity change of a neutron by measuring its Larmor phase $\Phi = 4 \pi f t$, where $t$ is the time it takes for the neutron to travel between the rf flippers, a difference in time is measured as a change in the neutron velocity. An uncorrected echo measurement will then conflate a change in scattering angle with a change in energy. The time can be written in terms of the neutron path length between the rf flippers as
\begin{equation}
     t = \frac{L_{\mathrm{RF}}}{v \cos\theta},
\end{equation}
where $v$ is the neutron velocity and $\theta$ is the scattering angle in the scattering plane (see Fig. \ref{figSchematic}).\cite{Keller2002}
Thus, for our example NRSE beamline, a neutron scattering at 1 degree would be out of Larmor phase by more than 2000 degrees compared to the unscattered neutrons if there were no correction.
This aberration will be present for elastic, quasielastic, and inelastic neutron scattering.
Clearly, an NRSE instrument must have a method for correcting this geometric contribution to the Larmor phase. In NSE instruments, Fresnel coils with longitudinal-fields (i.e. the field is orientated along the optical axis) are used for an analogous correction as well as correcting the aberration from the static field variation. \cite{Ohl2005,Ohl2012,Farago2015} However, transverse-field rf flippers (i.e. the rf flipper's static-field is perpendicular to the optical axis) have been constructed for NRSE measurements, \cite{Li2020, Endo2019} which require a transverse-field correction magnet to improve the polarization.

In this paper, we present an analytical solution of the magnetic field profile needed to correct the path length aberrations and the design of a suitable prototype correction magnet. The Larmor phase that a neutron acquires traveling through our prototype NRSE correction magnet varies quadratically with the distance from the optical axis and is radially symmetric. We simulate an NRSE beamline with the correction magnets and experimentally measure the spatial dependence of the Larmor phase change through the device. We demonstrate that transverse-field NRSE beamlines can be corrected with transverse static-field magnets, and therefore have the potential to be competitive with NSE beamlines.

\section{Analytical Solution} \label{sec:ana}

The two arms of the NRSE instrument will be corrected independently; we will discuss the correction for the second arm first.
The necessary magnitude of the correction is proportional to the path~length~difference~$\Delta L_{\mathrm{RF}}$ between the two pairs of rf flippers, which is given by
\begin{equation} \label{pathlength}
    \Delta L_{\mathrm{RF}} = L_{\mathrm{RF}} \left(\frac{1}{\cos\theta} - 1\right) \approx \frac{1}{2}L_{\mathrm{RF}}\theta^2,
\end{equation}
where $\theta$ is the small scattering angle shown in Fig. \ref{figSchematic}, defined relative to the optical axis.
This difference in path length will cause a delay in time, and thus a difference in the Larmor phase $\Phi$.
The difference in Larmor phase $\Delta \Phi$ between the unscattered and scattered beam for idealized rf flippers in the NRSE configuration is\cite{Keller2002}
\begin{equation} \label{deltaLinital}
    \Delta \Phi = \frac{4 \pi f}{v} \Delta L_{\mathrm{RF}} \approx 2 \pi f \frac{\lambda m}{h} L_{\mathrm{RF}} \theta^2.
\end{equation}
The Larmor phase has already been defined as $\Phi  = 4 \pi f t$ for an NRSE instrument, but it can also be defined for NSE instruments in terms of the magnetic field integral: $\Phi = (\gamma/v) \int ds \, B$, where $\mathrm{FI}_s = \int ds \, B$ the field integral experienced by the neutron traveling along the path $s$ and $\gamma \approx -1.832 \times 10^8$ rad/(T$\cdot$s) is the neutron's gyromagnetic ratio. 
In NSE, the neutron magnetic moment rotates (precesses) in the plane perpendicular to an applied static field and the Larmor phase measures the amount of this rotation relative to some fixed direction in the lab. In NRSE, the rf field rotates in the plane perpendicular to a static field, and the Larmor phase measures that angle between the neutron magnetic moment and the rf field.
Hence, rf and static field effects on the neutron can be added together, as has been exploited recently.\cite{Jochum2020b}

To correct the phase difference, we must design a correction scheme consisting of static magnetic fields that generates a Larmor phase proportional to $\theta^2$, with the proportionality coefficient $\chi$ being
\begin{equation} \label{CC}
   \chi  = 2 \pi \frac{f}{\gamma} L_{\mathrm{RF}}.
\end{equation}
Notice that the required correction is independent of wavelength. With this correction, the Larmor phase of all diverging 
neutrons will be corrected as if they traveled the same effective path length, namely the distance $L_{\mathrm{RF}}$.

Unfortunately, it is not obvious how to design a static magnetic field profile that would generate a purely $\theta^2$-dependent field integral term for a finite-sized beam.
However, we can generate such a term for a finite-sized beam with three correction magnets consisting of transverse static fields in which the magnitude of the field integral of a neutron traveling through the devices varies quadratically as a function of transverse position, with the center being the minimum and increasing radially outward. For scattering from a point-like sample, one can show that only two devices of this type are needed. 
This design is a two-dimensional extension to the original solution proposed by Monkenbusch. \cite{Monkenbusch1999}

For simplicity, we only look at the aberrations in one dimension (transverse $y$ direction), but the following argument can be easily generalized to the entire two-dimensional plane perpendicular to the optical axis. The neutrons may be scattered from any position on the sample at a distance of $y_{\mathrm{sam}}$ from the optical axis into any angle $\theta$ defined relative to the optical axis.
If the sample were point-like, then the scattering angle $\theta$ would be defined just by the $y$ distance from the optical axis and the $\theta^2$ aberration would be known simply from the $x$ and $y$ position in the second arm.  With a finite-size sample, the scattering position $y_{\mathrm{sam}}$ will add to the $y$ position from the scattering angle, so one device at a specific point along the beamline is not sufficient to correct the $\theta^2$ aberration.

To lowest order in scattering angle, the field integral per amp of a single prototype correction magnet is
\begin{equation} \label{correctiondeviceFIgeneral}
   \mathrm{FI}_n = a_n + b_n y_n^2 + c_n y_n \theta,
\end{equation}
where $y_n$ is the distance between the neutron's path at the $n^{\mathrm{th}}$ correction magnet and the optical axis (see Fig. \ref{figSchematic}). There is no linear term in $y$ because the device is left-right symmetric; similarly, there is no linear term in $z$ due to the top-bottom symmetry. Each term in the field integral is proportional to the applied current, and $a_n$ has units of T$\cdot$m, $b_n$ has units of T/m, and $c_n$ has units of T/rad. Here $b_n$ is the term that corrects the path-length aberrations while the $a_n$ and $c_n$ terms appear because of the particular correction magnet geometry that we have chosen; higher order terms were found to have a negligible contribution to the field integral.

The transverse position that the scattered neutron passes through each correction magnet is given by
\begin{equation} \label{ydefinition}
    y_n = y_{\mathrm{sam}}+ L_n \theta,
\end{equation}
where $\theta$ is assumed to be small and $L_n$ is the distance from the sample to the center of the $n$\textsuperscript{th} correction magnet, as shown in Fig. \ref{figSchematic}.
Combining Eqns. \eqref{correctiondeviceFIgeneral} and \eqref{ydefinition} for each device, we find the total field integral per amp $\mathrm{FI_T}$ experienced by a neutron arriving at the analyzer due to the three correction magnets to be
\begin{equation} \label{eqnLarmorPhaseCC}
\begin{aligned}
    \mathrm{FI_T} = \sum_{n \in \{4,5,6\}} \big[& a_n + b_n y_{\mathrm{sam}}^2 + (c_n + 2 b_n L_n)  y_{\mathrm{sam}} \theta \\
    &+ L_n (c_n + b_n L_n) \theta^2 \big].
\end{aligned}
\end{equation}
%
The goal of the correction scheme is to have the coefficient of the $\theta^2$ term equal to Eqn. \eqref{CC} while having all other terms zero. Doing so, the series of correction magnets would correct the path-length aberration in the second arm of the instrument regardless of the  $y_{\mathrm{sam}}$ position and without introducing any net field integral. These requirements on Eqn. \eqref{eqnLarmorPhaseCC} can be rewritten into several conditions:
\begin{gather*} \label{CoefficientCancellingEqns}
    a_4 + a_5 + a_6 = b_4 + b_5 + b_6 = c_4 + c_5 + c_6 = 0 \\
    b_4 L_4 + b_5 L_5 + b_6 L_6 = 0 \\
    \chi = L_4 (c_4 + b_4 L_4) + L_5 (c_5 + b_5 L_5) + L_6 (c_6 + b_6 L_6).
\end{gather*} 
Notice that if the sum of the currents through the three correction magnets is zero, then the first line will be satisfied.
Ignoring the $a_n$ terms for now, we solve this system of equations, obtaining 
\begin{subequations} \label{SolutionEnc}
\begin{align} 
    b_4 &= \frac{\chi + c_4(L_5 - L_4) + c_6(L_5 - L_6)}{(L_4 - L_5)(L_4 - L_6)} \\
    b_5 &= -\frac{\chi + c_4(L_5 - L_4) + c_6(L_5 - L_6)}{(L_4 - L_5)(L_5 - L_6)} \\
    b_6 &= \frac{\chi + c_4(L_5 - L_4) + c_6(L_5 - L_6)}{(L_4 - L_6)(L_5 - L_6)} \\
    c_5 &= - (c_4 + c_6),
\end{align}
\end{subequations}
with $c_4$ and $c_6$ being free parameters.
From this set of solutions, it is apparent that if we choose CM5, the fifth correction magnet, to be equidistant from CM4 and CM6 (such that $|L_4 - L_5| = |L_5 - L_6| = \delta L$) and also $c_4 = c_6$, then the angle-dependent $c_n$ terms cancel out, leaving $b_4 = b_6 = \chi/[2(\delta L)^2]$ and $b_5 = -\chi/(\delta L)^2$. Therefore, we can obtain our desired field integral by putting the same field in the first and last device and a field twice as large in the opposite direction in the middle device. We call this choice of fields the $(1,-2,1)$ configuration. In this configuration, the constant $a_n$ terms will also cancel out.

Next, we determine the magnitude required for $b_n$ for a realistic NRSE beamline. Plugging in Eqn. \eqref{CC} to the $b_n$ terms in Eqn. \eqref{SolutionEnc}, we see that
\begin{gather}
    b_4 = \frac{\pi f L_{\mathrm{RF}}}{\gamma (\delta L)^2} \\
    b_5 = -2 b_4, \quad b_6 = b_4. \nonumber
\end{gather}
To estimate the required field in the correction magnet, let $L_{\mathrm{RF}} = 2$ m, $\delta L = 1$ m, and $f = 4 $ MHz. Plugging in the numbers, we find $b_4 \approx -140$ mT/m. As we will show below, this value is attainable with our correction magnet.

The above discussion has only considered the second arm; now we look at the first arm. 
If the initial beam is well-collimated (i.e. all neutrons in the first arm travel parallel to the optical axis), then no correction elements are required in the first arm even if there are correction magnets in the second arm. However, in practice, neutrons in a real instrument have some divergence angle relative to the optical axis.
This initial beam divergence will lead to a variation in path length between neutrons propagating at different angles in the first arm, similar to the scattering term for the second arm.
Therefore, we must correct for this variation in the first arm with another three correction magnets, as shown in Fig. \ref{figSchematic}. They must be in the $(-1,2,-1)$ configuration as the static magnetic field in both rf flippers is in the opposite direction relative the static fields in the rf flippers in the second arm.
Without this additional correction, we do not obtain the best possible improvement to the polarization.
The correction for the divergence angle of the neutron in the first arm will not require any changes to the correction magnet set-up in the second arm because the correction for the second arm is independent of angle or $y_{\mathrm{sam}}$ position.

With all six correction magnets installed, the Larmor phase, and hence Fourier time, of neutrons along any path in either arm will be corrected to the Larmor phase and Fourier time of a neutron traveling parallel to the optical axis.

\section{Development of the Correction Magnet} \label{sec:design}

\begin{figure}
\includegraphics[width=.95\linewidth]{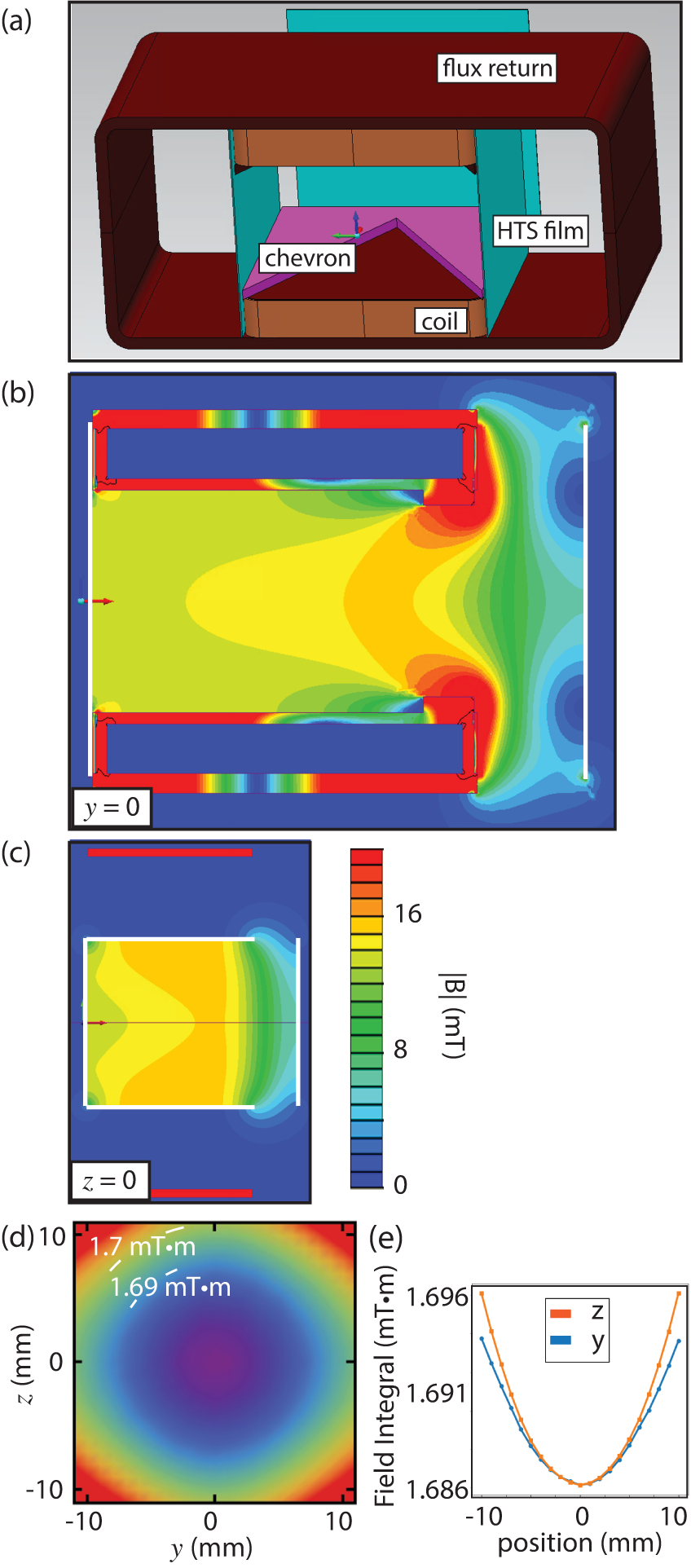}
\caption{\label{figCorrDevelopment} (a) CAD model of the correction magnet. The light blue surfaces are the high-temperature superconducting (HTS) films (front film not shown), the purple is the low-carbon steel chevron, the tan are the HTS coils, and the brown is the low-carbon steel flux return and pole pieces. (b and c) Simulation of fields at $y = 0$ and $z = 0$ with the coil current set to 10 amps, which have the same colorbar legend. The beam traveled from left to right. The HTS films are highlighted in white. (d) Contour plot of simulated field integral through the correction device at 10 amps for a neutron originating from a point source at $y = z = 0$ and $x = -20$ m. (e) Slices of the simulated field integrals in (d) through the origin.}
\end{figure}

To implement the analytical solution, we designed and constructed a vertical-field correction magnet following the drawing of Fig. \ref{figCorrDevelopment}(a). The bottom, top, and sides of the coils are enclosed by a magnetic flux return made of low-carbon steel (alloy 1018). High-temperature superconducting (HTS) films, gold-coated 350 nm thick YBCO on 0.5 mm thick sapphire substrate, were placed on the front and sides of the coils with another film placed 38 mm after the coils, outside of the magnetic circuit. These HTS films act as magnetic field screens due to the Meissner effect which creates sharp boundaries between field regions, as shown in Fig. \ref{figCorrDevelopment}(b,c).
Thus the magnet can be thought of in two parts: a contained region where the coils sit and an open region before the back film. HTS wire was wound around hollow low-carbon steel pole pieces and topped with ``chevrons'', low-carbon steel plates with v-shaped cutouts. The opening of the v is at the front of the device. The thickness of the chevrons was 3.2 mm, leaving a separation between chevrons of 50 mm. The angle of the chevron was 60 degrees, and the space from the coils to the rear HTS film was 38 mm. It was already well-known that a dipole magnet without a HTS film constraining the magnetic flux will create a magnetic field with a quadratic $z$-dependence, as correcting for this was one of the initial advantages of adding a HTS film. \cite{Wang2014}

The magnetic field in this device was simulated using the Siemens MagNet $\copyright$ software, which includes the material properties in its solutions via the finite-element method. The HTS films were simulated as perfect diamagnets preventing any perpendicular magnetic flux. We note that the explicit field profile in the device is arbitrary as long as the resulting field integral is quadratic across the device and the field direction does not change too quickly. 

A useful feature of this design is that the $y$ and $z$ components of the field integral may be tuned independently. The $z$ component, shown in Fig. \ref{figCorrDevelopment}(b) is largely dependent on the distance to the back film while the $y$ component, shown in Fig. \ref{figCorrDevelopment}(c) is largely dependent on the chevron angle. As shown in \ref{figCorrDevelopment}(b), the quadratic behavior of the $z$ component comes from the bowing field lines protruding around the back of the coils due to the displaced back film. 

A numerical solution to the field integral for any starting position $(y,z)$ and angle through the device was found by extracting the MagNet solution for the field at each point (mesh size 4 mm) and integrating the field along any chosen path. The simulated field integral through the coils at 10 amps for a neutron traveling from a far-away point source at $y= z =0$ and $x$ = -20 m is shown in Fig. \ref{figCorrDevelopment}(d). The difference in field integral between the center and edges is about 0.01 mT$\cdot$m, which approximately is the necessary correction value for an NRSE beamline with an rf flipper frequency of 1 MHz and 1 meter between the correction magnets. The field integral through the center is about 1.69 mT$\cdot$m.

\section{McStas Simulations of an NRSE Instrument} \label{sec:mcstas}
\FloatBarrier

Using McStas, a Monte Carlo neutron ray-tracing software package,\cite{Willendrup_Lefmann_2020, Willendrup_Lefmann_2021} we simulated an NRSE beamline with these correction magnets installed.  The polarizer, rf flippers, analyzer, and detector were taken to be 100$\%$ efficient, while the correction magnet component was built using numerically simulated magnetic field data extracted from the MagNet simulations. The following model was used for the rf flipper at resonance:
\begin{equation}
    \Phi_f = 2 \pi f (2t_i + \Delta t) -\Phi_i,
\end{equation}
where $\Phi_f$ is the final Larmor phase after exiting the rf flipper, $f$ the rf frequency (2 MHz for these simulations), $t_i$ the time at which the neutron enters the flipper, $\Delta t$ the time spent inside the flipper of thickness 15 cm, and $\Phi_i$ the initial Larmor phase when entering the flipper.
We used a modified version of the ``SANS\_spheres2'' sample, a default McStas sample that emulates elastically scattering hard spheres in a dilute solution. The sample parameters were chosen to prevent incoherent scattering or transmission without scattering, so the component acted like an idealized elastic scatterer with a maximum momentum transfer of 0.09 nm$^{-1}$, which corresponds to a maximum scattering angle of 0.7 degrees. The sample was 3 cm by 3 cm in transverse size with negligible thickness. The aperture diameter and the neutron wavelength were 2 cm and $0.8$ nm $ \pm 1\%$, respectively.
The separation between the rf flippers in each arm was taken to be 2.3 m, and the distance between the correction magnets 1 meter. The total distance from the source to the two-dimensional detector was 5.2 meters. With these parameters, the effective initial beam divergence is about 0.6 degrees.

Using the magnetic field data extracted from MagNet simulations, we determined that our prototype correction magnet had the following field integral per amp expansion across the correction magnet:
\begin{equation} \label{Eqn:fullFIE}
    \mathrm{FI} = a + b_y y^2 + b_z z^2 + c_y y \theta + c_z z \psi,
\end{equation}
where $\psi$ is the vertical neutron divergence angle and the values of the fitted coefficients are given in Tab. \ref{tab:FIE CC coefficients} below. 
Higher order terms were found to have a negligible contribution.

\renewcommand{\arraystretch}{2}
\begin{table}[h]
\centering
\newcolumntype{R}{>{\centering\arraybackslash}X}
\begin{tabularx}{.99\linewidth}{R|R|R|R|R}
       $a \left(\frac{\mathrm{mT}\cdot \mathrm{m}}{\mathrm{A}}\right)$ &
       $b_y \left(\frac{\mathrm{mT}}{\mathrm{A}\cdot\mathrm{m}}\right)$ &
       $b_z \left(\frac{\mathrm{mT}}{\mathrm{A}\cdot\mathrm{m}}\right)$ &
       $c_y \left(\frac{\mathrm{mT}}{\mathrm{A}\cdot\mathrm{rad}}\right)$ &
       $c_z \left(\frac{\mathrm{mT}}{\mathrm{A}\cdot\mathrm{rad}}\right)$ \\
\hline
 0.169 & 8.78 & 8.97 & -0.525 & 0.986
\end{tabularx}
\renewcommand{\arraystretch}{1}
\caption{\label{tab:FIE CC coefficients} 
Values of the field integral expansion coefficients in Eqn. \eqref{Eqn:fullFIE}. The longitudinal length of the magnet is 14 cm.}
\end{table}

\begin{figure}[!b]
\centering
\includegraphics[width=.95\linewidth]{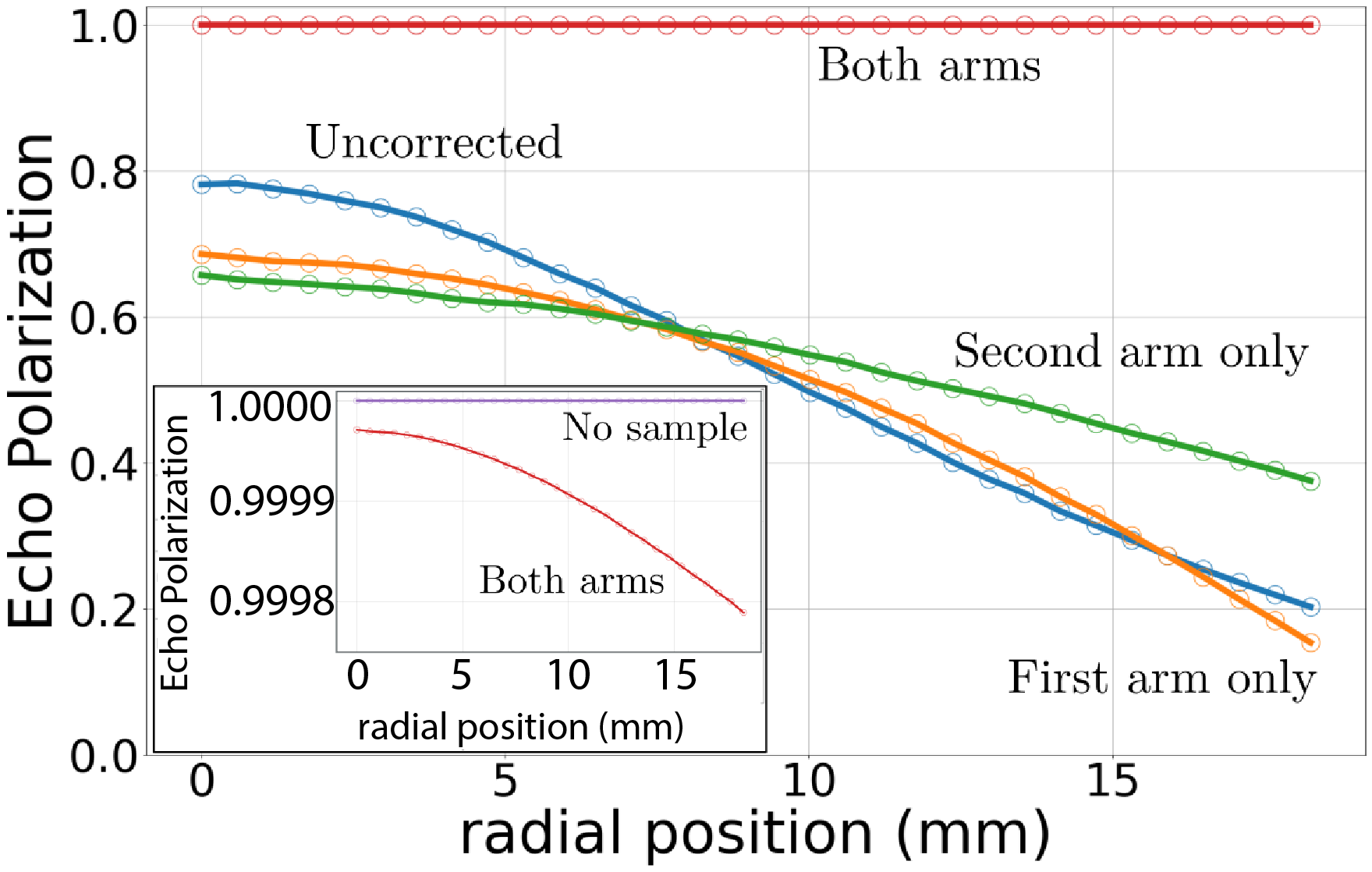}
\caption{\label{figNRSEmcstas} Simulated McStas polarization for an NRSE beamline with all six correction magnets on (red curve), only the final three correction magnets on (green curve), only the first three correction magnets on (orange curve), and no correction magnets on (blue curve).
The inset compares the polarization for all correction magnets on (red curve) to all correction magnets off and sample removed (purple curve) which shows that the polarization drops to about 0.9998 at the edge of the detector when both arms are corrected.}
\end{figure}

The NRSE beamline was simulated in McStas both with and without the correction magnets.
A plot comparing the simulated echo polarizations vs. radial position on the detector is shown in Fig. \ref{figNRSEmcstas}.
These simulations confirm that the inclusion of the correction elements greatly increases the polarization, especially for larger scattering angles which correspond to the edges of the detector.
Correcting only the second arm of the beamline improves the polarization for the larger scattering angles, although the polarization at the center of the detector is worsened compared to the uncorrected simulation due to the initial neutron divergence angle. However, if the initial beam divergence is large (e.g., about 1 degree or more for our specific simulation parameters), then both of the single arm correction schemes show very little improvement in the polarization, so both arms must be corrected. The alignment of the correction magnets is also important, with more precision required for higher Fourier times. For the simulation parameters used above, all correction magnets must be aligned within approximately $\pm 0.5$ mm in both the $y$ and $z$ directions.

\section{Experiment Results} \label{sec:experiment}

\begin{figure}[b]
\centering
\includegraphics[width=.95\linewidth]{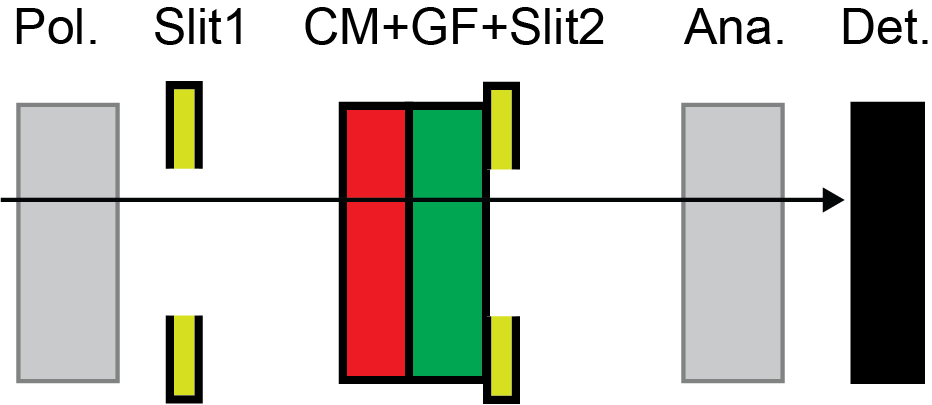}
\caption{\label{figBeamline} Schematic of the experimental test of the correction magnet (CM). The beam travels from left to right. Precession occurred inside both the CM and guide field (GF).}
\end{figure}

\begin{figure*}[t]
\centering
\includegraphics[width=.95\textwidth]{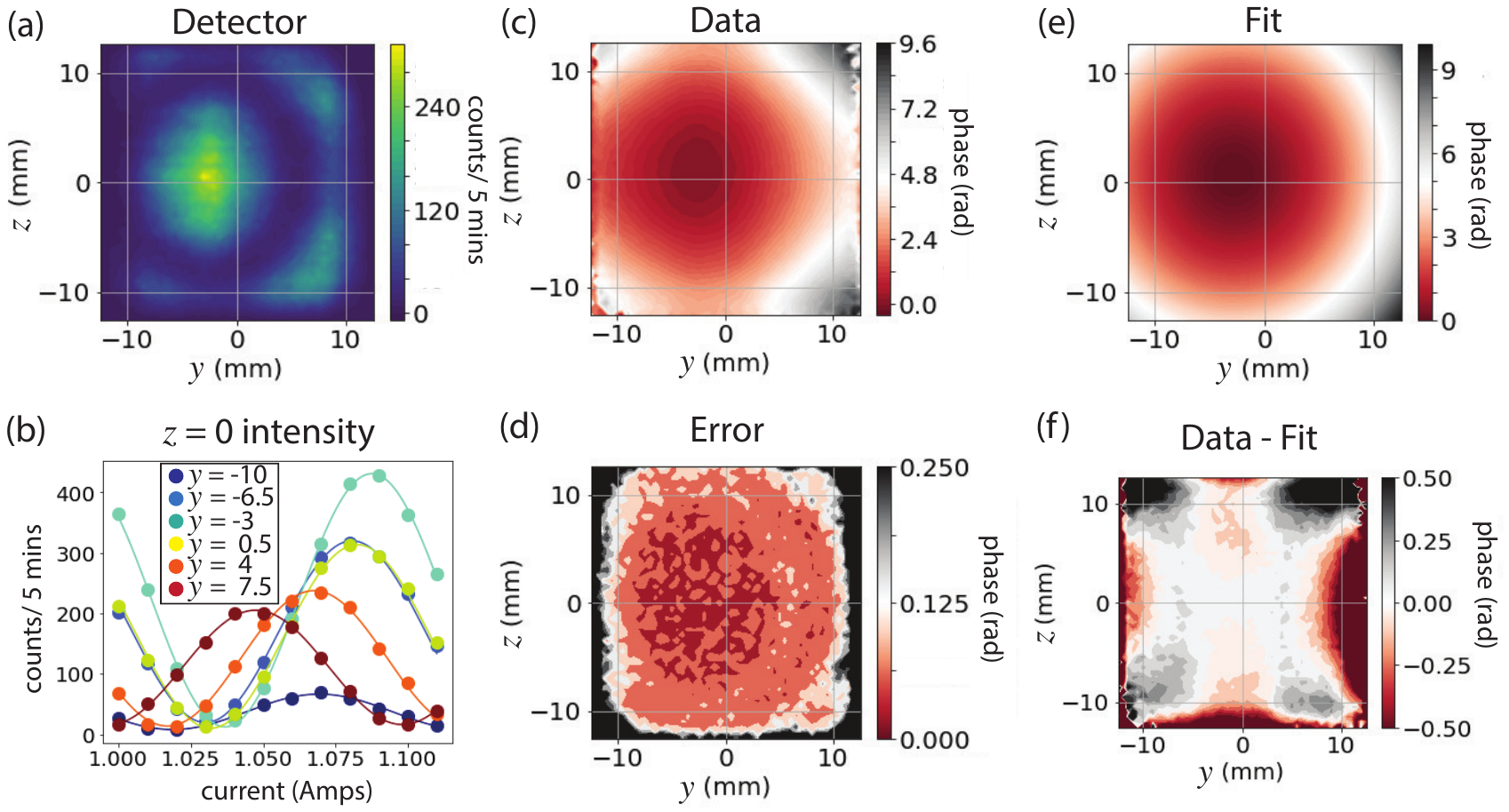}
\caption{\label{figCorrAnalysis} Data with a current of -10 amps in the NRSE correction magnet. (a) The intensity vs. position recorded by the Anger camera when the guide field coil had a current of 1.11 amps. (b) A cosine fit of the intensity vs. guide field coil current for several pixels along the line $z=0$. (c) The phase and (d) phase error extracted from the cosine fit shown for each pixel. (e) The quadratic fit of the phase data to Eqn. \eqref{Quadratic2d} and (f) the quadratic fit subtracted from the phase data. }
\end{figure*}

\begin{figure}
\centering
\includegraphics[width=.95\linewidth]{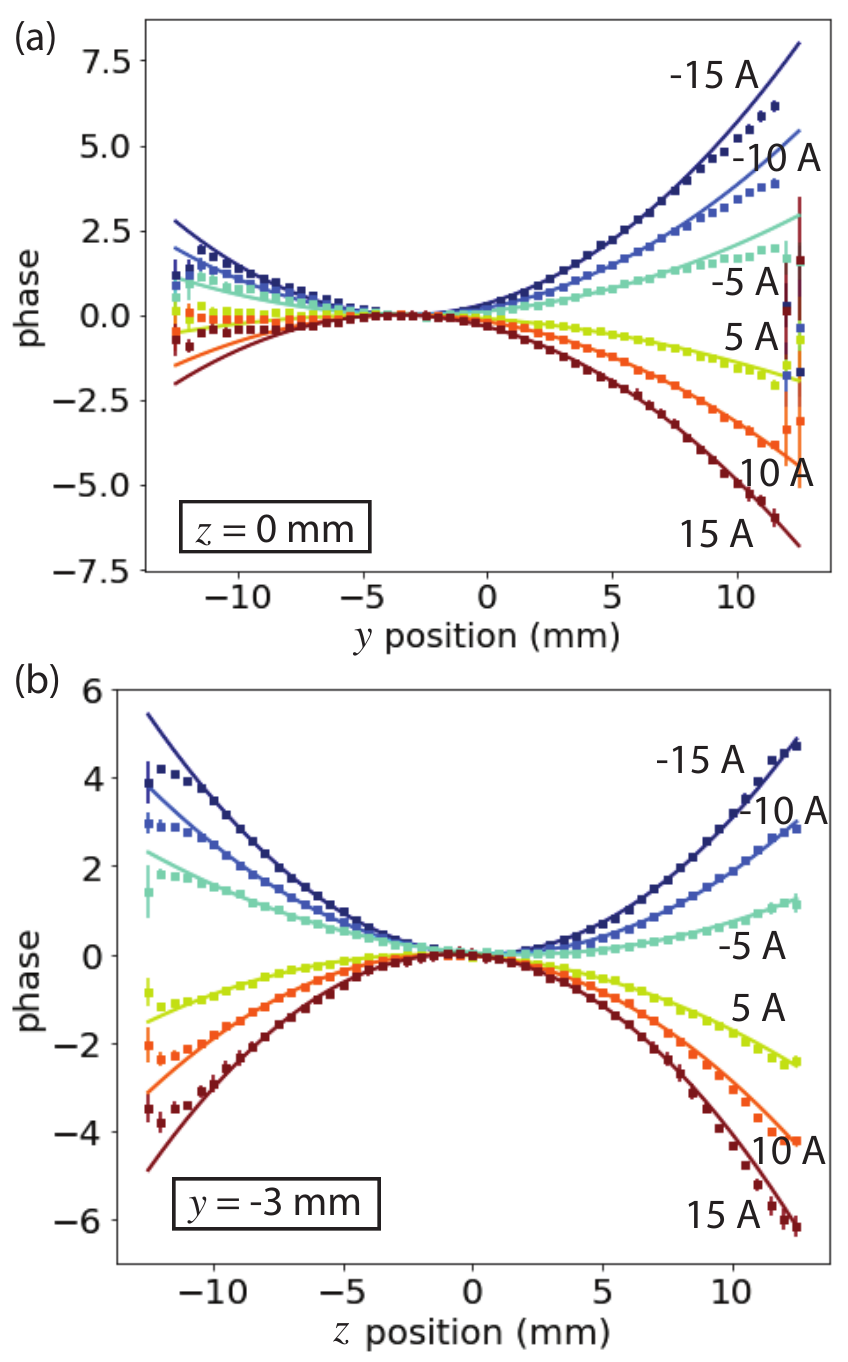}
\caption{\label{figCorrAnalysis1d} (a) Horizontal and (b) vertical slices through the magnetic center for different currents in the correction magnet. Data are fit to a parabola.}
\end{figure}

A measurement of the field integral through the prototype correction magnet was performed on the cold-neutron, polarized test beamline CG4B at the High Flux Isotope Reactor (HFIR) at Oak Ridge National Laboratory (ORNL). As shown in Fig. \ref{figBeamline}, the correction magnet was installed in a vacuum chamber in front of a guide field magnet that also generated a field in the vertical, $z$-direction. The guide field magnet had a front and back HTS film, with the front film also serving as the back film of the correction magnet. The vertical guide field magnitude was designed to be spatially uniform, as the neutron will continue to precess in it.
S-benders served as the neutron polarizer and analyzer. A horizontal guide field outside of the correction magnet and the non-adiabatic field transition through the HTS film induced precession inside both the correction magnet and the spatially uniform guide field. Precession was stopped by another horizontal guide field after the back HTS film of the guide field. The beam size was determined by a square 1 by 1 cm slit located 1 meter in front of the correction magnet and a square 2.5 by 2.5 cm slit at the end of the guide field coil. The wavelength was 0.55 nm with a FWHM wavelength spread of less than 1$\%$.

The variation in the Larmor phase across the beam was measured across the two-dimensional detector. The detector was an Anger camera with 1.8 mm pixel size, as shown in Fig. \ref{figCorrAnalysis}(a). \cite{Riedel2015, Cao2018} From the detector image in Fig. \ref{figCorrAnalysis}(a), one can directly see an approximately ``bullseye'' shaped signal, suggesting a radial dependence of the Larmor phase.
The Larmor phase was measured by setting a current in the correction magnet and scanning the precessing guide field between 1 and 1.12 amps. This current range varies the phase of neutrons passing through the magnet by about $2 \pi$, as shown in Fig. \ref{figCorrAnalysis}(b). The difference in the phase of the curves shown in Fig. \ref{figCorrAnalysis}(b) shows the different Larmor phase acquired by neutrons traveling through the correction magnet. The intensity recorded in each pixel varies greatly due to the spatial non-uniformity in the CG4B beam intensity as well as the non-uniform detector efficiency.

The intensity $N$ as a function of pixel $(y,z)$ was fit to 
\begin{equation}
N(y,z)  = \alpha + \beta \cos[\phi + f_g(I - I_0)],
\end{equation}
where $\alpha$ and $\beta$ are fitting parameters, $\phi$ is the Larmor phase from the correction magnet, $f_g$ is the frequency of the oscillation in the polarization due to the Larmor phase produced by the guide field, $I$ is the current in the guide field, and $I_0$ is the current at the start of the scan (1 amp in this case).
The polarization of the signal is defined as $\beta/\alpha$. We fit the relative phase compared to the center which we set as zero.

The phase data for the phase $\phi$ were fit to the following two-dimensional quadratic function:
\begin{equation} \label{Quadratic2d}
    \phi(y,z) = \phi_2 [(y - y_0)^2 + \epsilon(z-z_0)^2] + \phi_0,
\end{equation}
where $\phi_2$ and $\phi_0$ are fitting parameters and $\epsilon$ the eccentricity term which allows for a difference in the $y$ and $z$ correction terms, and $(y_0, z_0)$ the beam center. The fit is shown in Fig. \ref{figCorrAnalysis}(e). Subtracting the quadratic fit from the phase data gives the accessible corrected beam size, shown in Fig. \ref{figCorrAnalysis}(f) to be about 2 centimeters. MagNet simulations show that the eccentricity term can be tuned to unity by varying the chevron angle and back film separation distance.

There are several unexpected features in the data. Most notably, the center of the beam is not the same as the center of the quadratic fit, which we call the magnetic center. This discrepancy is possibly due to a misalignment of the beam mask and the correction magnet. Additionally, the positive $y$ side of the magnet has a larger discrepancy in the data-fit compared to the negative side. While these features are surprising, the most likely source for the off-center signal is due to misalignment between the beam apertures and the correction magnet, and the non-homogeneous signal is possibly due to a magnetic inhomogeneity in the soft-iron used in the pole and chevron pieces.
An additional feature is the non-radial dependence (i.e. the diamond shape of the fitted phase) of the data at large $(y,z)$. This feature can partially account for where the quadratic fit fails to match the data in Fig. \ref{figCorrAnalysis}(f). It can also be seen in the MagNet simulations of the field integral, suggesting that it is a result of the chevron design. A more sophisticated pole piece shape may be required to adjust the field integral into the proper quadratic shape for larger beam sizes. 

In order to compare different currents in the correction magnet, we fit a vertical and horizontal slice through the magnetic center to a parabola as displayed in Fig. \ref{figCorrAnalysis1d}. The offset in $y$ of 3 mm remains approximately constant for all currents, which is consistent with the conclusion of misalignment between the correction magnet and the beam. The quadratic coefficient divided by the current should be the same for all currents if the field is generated solely by the current in the correction magnets. However the phase change at 15 amps is only 2.6 times the variation at 5 amps. This difference is possibly due to hysteresis effects and domain formation in the soft-iron inside of the correction magnet. It may also be due to the coupling of the field in the correction magnet to the external guide fields, although MagNet simulations show very little coupling. 

\section{Discussion and Conclusion} \label{sec:dis}

This correction technique is for transverse-field NRSE instruments, while Fresnel coils may be installed for longitudinal NRSE.
An advantage of this device compared to Fresnel coils is the small amount of neutron-absorbing or scattering material in the beam. There is a 100 nm film of gold coating a 350 nm film of YBCO on a 0.5 mm sapphire substrate. With a 1 nm neutron wavelength, 12 of these films will have a transmission of $\sim 95\%$. Fresnel coils add at least 2 cm of aluminum wire which has a transmission of $\sim 86\%$ for 1 nm neutron wavelength. The exact amount of material for a Fresnel coil depends on the required current, so reaching a higher Fourier time generally produces more background scattering. However, the Fresnel coils have a long history of being successfully used to correct for NSE and have been built to accommodate much larger beam sizes. 

One of the reasons longitudinal NRSE is preferred for the Reseda instrument at FRM-II is the historical difficulty in correcting transverse NRSE path length aberrations. \cite{Franz2019} If this correction technique can reach the same performance as Fresnel coils, the choice of longitudinal or transverse-NRSE will be more complicated: transverse rf flippers offer the opportunity to have a higher effective frequency in ``bootstrap'' mode, \cite{Li2020} while longitudinal rf flippers have a proven history of high performance.\cite{Franz2019}

Another method of correcting divergent neutrons has recently been installed at VIN-ROSE in JPARC, which addresses the same problem by adding elliptical mirrors to each arm so that all neutron paths will be the same length. \cite{Endo2019} To our knowledge, this correction magnet has not yet been used to correct for scattering from different points in the sample.

We have demonstrated a theory, simulation, and experiment of correcting aberrations caused from path deviations in a transverse NRSE beamline.
Simulation shows that an arrangement of six correction magnets maintains a high polarization even for 3 cm beam sizes and large rf flipper frequencies.
Experimental tests of the prototype magnet show that shaping the pole pieces and separating the coil from the back HTS film allow for the independent control of the $y$ and $z$ parameters of a quadratic field integral.
The most pressing improvements to future designs are more careful alignment, more magnetically-uniform material for pole pieces, and acceptance of larger beam sizes.
With these improvements, this type of correction magnet is ready to benefit transverse NRSE beamlines.

\section{Acknowledgements}
The authors would like to thank Lowell Crow, Georg Ehlers, Fumiaki Funama, and Steven Parnell for useful discussions. CAD drawings were made by Jak Doskow and machining was done by the Indiana University Physics machine shop: John Frye, Danny Clark, Darren Nevitt, and Todd Sampson. We thank Matthew Loyd for assistance with the Anger camera.

This research used resources at the High Flux Isotope Reactor, a DOE Office of Science User Facility operated by the Oak Ridge National Laboratory. F. Li would also like to acknowledge the support from DOE Early Career Research Program Award (KC0402010), under Contract No. DE-AC05- 00OR22725.

The work reported here was funded by the Department of Energy STTR program under grants DE-SC0021482 and DE-SC0018453.
A number of the authors acknowledge support from the US Department of Commerce through co- operative agreement number 70NANB15H259.


\bibliographystyle{apsrev4-2.bst}
\bibliography{NRSEcorrbib.bib}

\section{Supplemental Material} \label{sec:appx}

The following table contains the fitting parameters for the HFIR experiment.

\begin{figure*}
\centering
\includegraphics[width=.95\textwidth]{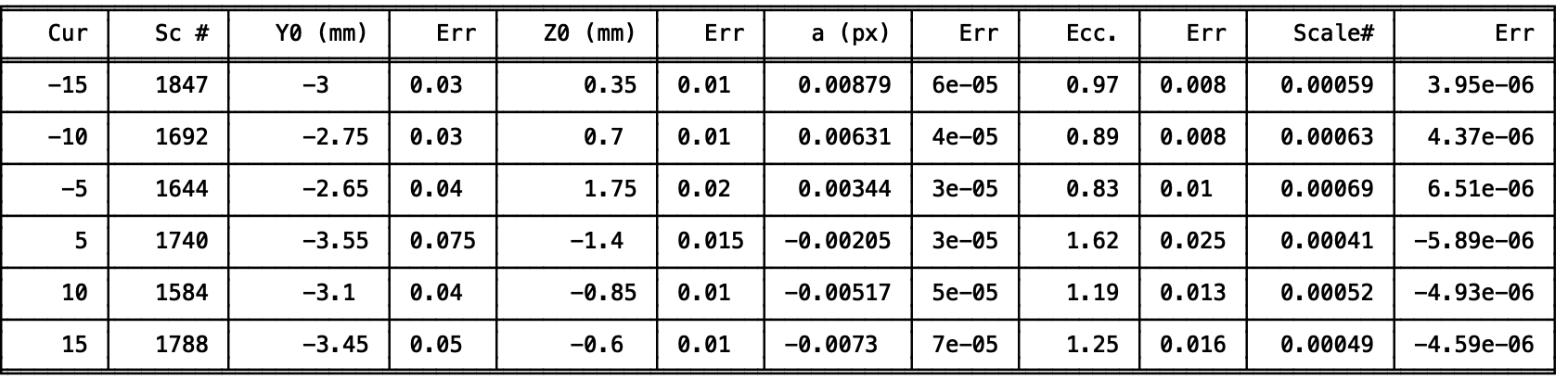}
\caption{\label{figtable} Table of fitting parameters for the HFIR experiment.}
\end{figure*}

The following plot shows the fitted phase, phase error, and fitted polarization for various correction coil currents.

\begin{figure*}[p]
\centering
\includegraphics[width=.85\textwidth]{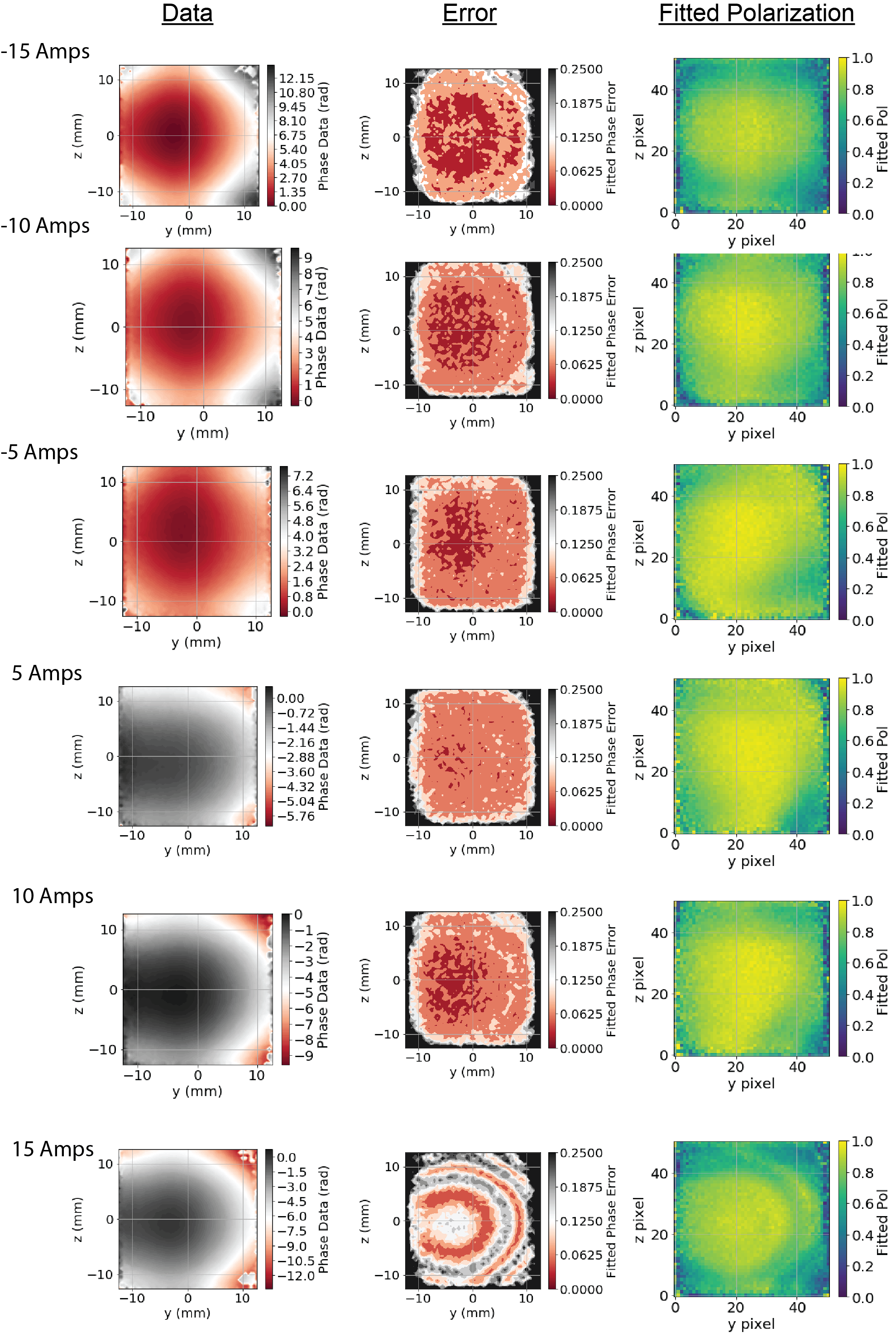}
\caption{ \label{figAllData} Experimental data for the HFIR experiment.}
\end{figure*}




\end{document}